\documentstyle[aps,prb,epsfig,preprint,tighten,floats,eqsecnum]{revtex}

\def\s{\sum\limits}

\def\pa{\partial}
\def\i{\int\limits}
\def\be{\begin{equation}}
\def\e{\end{equation}}
\def\beml{\begin{mathletters}}
\def\eml{\end{mathletters}}
\def\beq{\begin{eqnarray}}
\def\eq{\end{eqnarray}}
\def\ba{\begin{array}}
\def\a{\end{array}}

\def\l{\left}
\def\r{\right}
\def\n{\nonumber}
\def\la{\langle}
\def\ra{\rangle}

\def\im{\,{\rm Im}\,}

\def\ep{\varepsilon}

\begin{document}
 
\draft

\date{March 2000}

\title{Andreev levels in a single-channel conductor}

\author{M.\ Titov$^{1}$, N. A.\ Mortensen$^{1,2}$, H.\ Schomerus$^{1,3}$,
and C. W. J.\ Beenakker$^{1}$} 
\address{
{}$^1$Instituut-Lorentz, Universiteit Leiden,
P.\,O.~Box 9506, 2300 RA Leiden,
The Netherlands\\
{}$^2$Mikroelektronik Centret, Technical University of Denmark, \O rsteds
Plads 345 E, 2800 Lyngby, Denmark\\
{}$^3$Max-Planck-Institut f\"ur Physik komplexer Systeme, N\"othnitzer Str. 38,
01187 Dresden, Germany
}
\widetext

\maketitle

\begin{abstract}
We calculate the sub-gap density of states of a disordered single-channel normal metal
connected to a superconductor at one end (NS junction) or at both ends (SNS junction).
The probability distribution of the energy of a bound state (Andreev level) is broadened by
disorder. In the SNS case the two-fold degeneracy of the Andreev levels
is removed by disorder
leading to a splitting in addition to the broadening. 
The distribution of the splitting is given precisely by 
Wigner's surmise from random-matrix theory.
For strong disorder
the mean density of states is largely unaffected by the proximity to the superconductor,
because of localization, except in a narrow energy region near the Fermi level, where 
the density of states is suppressed with a log-normal tail.
\end{abstract}
\pacs{PACS numbers: 74.80.Fp, 72.15.Rn, 73.63.Rt}

\narrowtext

\section{Introduction}

Several recent works have identified and studied 
deviations from mean-field theory in the sub-gap density
of states of a normal metal 
in contact with a superconductor.\cite{Beloborodov,Simons,Vavilov,Skworzov}
The excitation spectrum below the gap of the bulk superconductor
consists of a coherent superposition of electron and hole excitations,
coupled by Andreev reflection\cite{Andreev} at the normal-metal--superconductor
(NS) interface. The energy of these Andreev levels fluctuates from sample to sample, but
such mesoscopic fluctuations are ignored in mean-field theory.
Because of these fluctuations, the ensemble averaged density
of states $\la \nu(\ep)\ra$ acquires a tail that extends
below the mean-field gap, vanishing only at the Fermi level
(zero excitation energy $\ep$). The fluctuations become particularly
large if the size of the normal metal is greater than the
localization length.

The purpose of this paper is to analyze an extreme case of complete 
breakdown of mean-field theory, which is still sufficiently 
simple that it can be solved exactly. This is the case of 
single-mode conduction through a disordered 
normal-metal wire attached to a superconductor.
The localization length in this geometry is equal to 
the elastic mean free path $\ell$, so that the wire crosses over
with increasing length $L$ from the ballistic regime directly
into the localized regime --- without an intermediate diffusive regime.
Perturbation theory is possible in the quasiballistic regime
$\ell \gg L$, but for $\ell < L$ an essentially non-perturbative approach is
required. We will use an approach based on a scaling equation (also known as invariant
embedding), that has proved its use before in different 
contexts.\cite{CWJreview,White,TB,TBMF} 

We will contrast the quasiballistic and localized regimes,
as well as the two geometries with a single superconducting 
contact (NS junction)  or with two superconducting contacts at both ends 
of the normal metal wire (SNS junction). If we assume that the 
two superconductors have the same phase, so that there is no 
supercurrent flowing through the normal metal, then the 
Andreev levels of the SNS junction are doubly degenerate 
in the absence of disorder. This degeneracy is broken
by disorder. We find that for weak disorder the probability 
distribution of the splitting is given precisely
by Wigner's surmise from random-matrix theory.\cite{Mehta}
(The spectra of
chaotic systems have spacings described by Gaudin's
distribution, which is close to, but not identical with
Wigner's surmise.\cite{Mehta})

In the localized regime the fluctuations of the Andreev levels become greater 
than their spacing, and they can no longer be distinguished in the mean density of
states, which decreases smoothly to zero on approaching the Fermi level.
The energy scale for this soft gap is exponentially small because of
localization, given by $\ep_g=(\hbar v_F/\ell)e^{-L/\ell}$.
The decay of $\la \nu(\ep)\ra$ for $\ep\ll \ep_g$ 
has a log-normal form $\propto \exp\l[-\case{\ell}{4L}
\ln^2\l(\ep_g/\ep \r)\r]$.
Such log-normal tails are characteristic of rare fluctuations 
in the localized regime\cite{Mirlin} and have appeared recently
in the context of the superconductor proximity effect.\cite{Skworzov}
\vfill

\section{Quasiballistic regime}

\subsection{NS junction}

\begin{figure}
\hspace*{3cm}
\epsfxsize10cm
\epsfbox{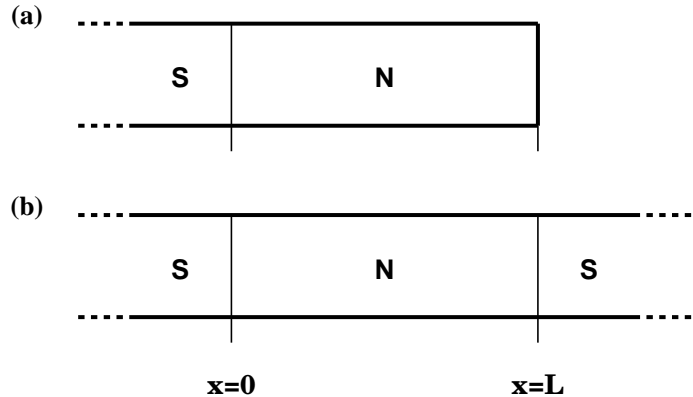}
\medskip\\
\caption{
Geometry of the NS and SNS junctions.
}
\label{fig:fig0}
\end{figure}

The NS junction consists of a piece of normal metal of length $L$
connected at one end to a superconductor (see Fig.\ {\ref{fig:fig0}}a).
The width of the normal metal is of the order of the 
Fermi wave length $\lambda_F$, such that there is a single propagating mode
at the Fermi energy $E_F$. We assume an ideal junction, without 
any tunnel barrier and with $E_F$ much greater than the superconducting gap 
$\Delta_0$.
An electron incident on the superconductor with energy $\ep<\Delta_0$
above the Fermi level is then Andreev reflected as a hole at energy
$\ep$ below the Fermi level,
with the phase shift 
\be 
\phi_A=-\arccos{(\ep/\Delta_0)}, \qquad -\pi/2 <\phi_A <0.
\e
We wish to know at which $\ep$ a bound state (Andreev level)
will form in the normal metal.

The electron and hole components of the wave function 
$\psi(x)=\biglb(u(x),v(x)\biglb)$ satisfy the Bogoliubov-de Gennes 
(BdG) equation\cite{deGennes}
\be
\l(\ba{cc} {\cal H}_0 & \Delta  \\
\Delta^* & -{\cal H}_0^*
\a \r) \psi = \ep \psi,
\e
where ${\cal H}_0=-(\hbar^2/2m)\pa^2/\pa x^2+V(x)$
is the Hamiltonian of the normal metal (with disorder potential $V$)
and $\Delta(x)=\Delta_0\theta(-x)$ is the superconducting gap
(which vanishes in the normal-metal region $x>0$). For narrow junctions
(width much less than the superconducting coherence length 
$\xi_0=\hbar v_F/\Delta_0$) the depletion of $\Delta(x)$
on the superconducting side may be neglected,
hence the step function $\theta(-x)$.
At the closed end $x=L$ of the normal metal 
we impose the boundary condition $\psi(L)=0$.

In this section we address the quasiballistic regime 
of mean free path $\ell \gg L$. We can then treat $V$ as a small perturbation
on the ballistic bound states
\widetext
\beml
\beq
\psi(x)&=&{ 1\over \sqrt{Z} }
\l(
\ba{c}
\sin[( k_F+k )(x-L)] 
\\
\sin[( k_F-k )(x-L)-\pi n]
\a 
\r),\qquad\qquad\qquad\qquad\quad 0<x<L, \\
\psi(x)&=&{ 1\over \sqrt{Z} }
\l(
\ba{c}
\sin[k_F x -(k_F+k) L] 
\\
\sin[k_F x -(k_F-k) L-\pi n] 
\a \r)\exp{\l(-\frac{x}{\xi_0}\sin{\phi_A}\r)},
\; \qquad x<0. 
\eq
\eml
\narrowtext\noindent
The normalisation constant is
$Z=L-\case{1}{2}\xi_0/\sin{\phi_A}$ 
for $k_F L \gg 1$.
(We denote $k_F=m v_F/\hbar=2\pi/\lambda_F$.)
The wave number $k=\ep/\hbar v_F$
should satisfy the quantization condition
\be
\label{NSlevels}
2 k L +\phi_A= \pi n, \qquad n=0,1,2,\dots .
\e
The total number of Andreev levels within the gap is
$2L/\pi \xi_0$ for $L\gg \xi_0$. (There remains one level if
$L\ll \xi_0$.)  

To first order in $V$ the energy level is shifted by the
matrix element 
\be
\label{shift}
\delta \ep=\int_0^L dx\, 
V(x) \l[ u(x)^2-v(x)^2 \r].
\e
We assume a potential with a short-range 
correlation, expressed by
\be
\label{potential}
\la V(x)\ra=0, \quad
\la V(x) V(x')\ra={\hbar^2 v_F^2 \over \ell}\delta(x-x'),
\e
where $\la \cdots \ra$ stands for the disorder average.
It follows that the distribution of an Andreev level 
around its ballistic value is a Gaussian with zero mean, 
$\la \delta\ep \ra=0$, and  variance
\be
\label{var}
\la \delta\ep^2 \ra=
{\hbar^2 v_F^2\l(2 L+ \xi_0 \sin\phi_A \r)\over
2\ell \l(2 L- \xi_0/\sin{\phi_A}\r)^2}.
\e

\begin{figure}
\hspace*{3cm}
\epsfxsize10cm
\epsfbox{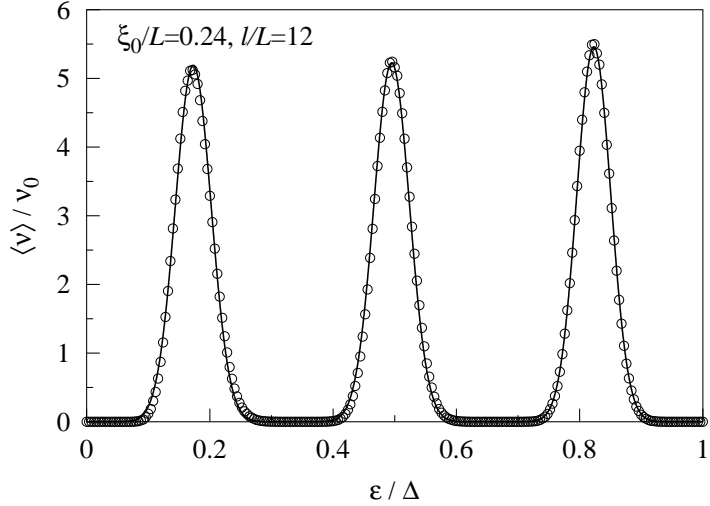}
\medskip\\
\caption{Mean density of states (in units
of $\nu_0=2L/\pi\hbar v_F$)
of a quasiballistic NS junction.
The Gaussian with variance given by 
Eq.\ (\ref{var}) (solid curves)
is compared to
the numerical solution of the 
BdG equation (data points).}
\label{fig:NS}
\end{figure}

By way of illustration, we show in Fig.\ \ref{fig:NS} the mean density of states
of an NS junction containing three Andreev levels ($\xi_0/L=0.24$)
with mean free path $\ell=12 L$. The Gaussian given by Eq.\ (\ref{var})
agrees very well with the numerical solution of the 
BdG equation (data points).

We briefly explain the numerical method.
The BdG equation is solved numerically on a one-dimensional grid
(lattice constant $a$) by replacing the Laplacian by finite differences
and truncating the Hamiltonian matrix in the superconducting
region, where the wave function is evanescent for energies in the
superconducting gap.
The resulting tight-binding model has nearest-neighbor coupling
$\gamma=\hbar^2/2m a^2$ (band width $4\gamma$).
We set $E_F=\gamma$ and $\Delta_0=0.1\,\gamma$,
corresponding to $\lambda_{\rm F}=6\,a$ and
$\xi_0=10\sqrt{3}\,a$.
The disorder is modelled by a random on-site potential
which is uniformly distributed in the interval $(-W,W)$.
The mean free path from the Born approximation,
$l=3 E_F(4\gamma-E_F) a/W^2$, was found to fit well
to the prediction of one-dimensional scaling theory for
the mean inverse transmission probability,
$\la T^{-1}\ra=\case{1}{2}[1+\exp(2 L/l)]$,
in the complete range from the quasiballistic to the localized regime.
(The localization length $\xi$ is related to the 
mean free path by $\xi=2 \ell$, cf. Ref.\ \onlinecite{CWJreview}.)
This allows for a parameter-free comparison of the analytical and numerical
results for the ensemble-averaged density of states.

\subsection{SNS junction}

The quasiballistic regime in an SNS junction (Fig.\ \ref{fig:fig0}b)
is qualitatively different from the NS case of the previous subsection.
The reason is the double degeneracy of the unperturbed Andreev 
levels.
This degeneracy exists if the phase of the order parameter in 
the two superconductors is the same, which is what we assume in
this paper. Let us examine the splitting of the Andreev levels
by the disorder potential.

The SNS junction has energy gap
\be
\Delta(x)=\Delta_0\theta(-x) +\Delta_0 \theta(x-L). 
\e
The quantization condition reads
\be
\label{SNSlevels}
k L+\phi_A=n \pi,\qquad n=0,1,2,\dots .
\e
There are $L/\pi\xi_0$ Andreev levels (for $L\gg \xi_0 $), 
each level being doubly degenerate.
We choose the two independent
eigenfunctions $\psi_{\pm}(x)$ such
that they carry zero current. 
They are given by
\widetext
\beml
\label{SNSfunc}
\beq
&& 
\psi_+(x)={1\over \sqrt{Z'}} 
\l(\ba{c}
\cos(k_F x)\\
\cos(k_F x-\phi_A)
\a \r)
\exp{\l(-\frac{x}{\xi_0}\sin\phi_A\r)},
\qquad\; x<0,
\\
&&\psi_+(x)={1\over \sqrt{Z'}}
\l(\ba{c}
\cos[(k_F+k)x] \\
\cos[(k_F-k)x - \phi_A] 
\a \r),\qquad\qquad\qquad\quad 0<x<L, 
\\
&&\psi_+(x)=
{1\over \sqrt{Z'}}
\l(\ba{c}
\cos( k_F x + k L)\\
\cos(k_F x+\pi n)
\a \r)\exp{\l( \frac{x-L}{\xi_0}\sin\phi_A \r)},
\qquad x>L,
\eq
\eml
\narrowtext\noindent
and $\psi_-(x)$ is obtained
by replacing cosine by sine.
The normalization constant
is now $Z'=L-\xi_0/\sin\phi_A$. 

To first order in $V$ the levels are splitted 
symmetrically around the ballistic value, by an amount
$\pm \case{1}{2}s$. The basis (\ref{SNSfunc})  is chosen in such a way 
that the off-diagonal elements of the 
perturbation vanish.
The shift of each level can then be calculated from
Eq.\ (\ref{shift}) using the corresponding eigenfunction.
We again 
calculate the probability distribution $P(s)$
of the level splitting using Eq.\ (\ref{potential}). The result is
\be
\label{surmise}
P(s)={\pi s\over 2 \la s \ra^2} \exp \l(-
{\pi s^2 \over  4 \la s\ra^2} \r),
\e
with average splitting
\be
\label{s:average}
\la s\ra=\Delta\sqrt{\pi\over 2\ell} {\xi_0 \sqrt{L+\xi_0\sin\phi_A}\over
L-\xi_0/ \sin \phi_A}.
\e
We recognize Eq.\ (\ref{surmise}) as Wigner's surmise of random matrix 
theory.\cite{Mehta}

\begin{figure}
\hspace*{3cm}
\epsfxsize10cm
\epsfbox{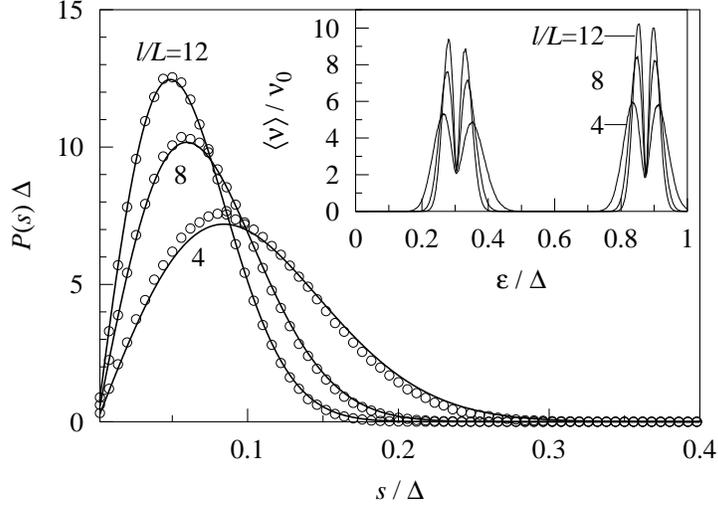}
\medskip\\
\caption{ 
Distribution of the
splitting $s$ of the first pair of
Andreev levels in an SNS junction with 
$\xi_0/L=0.24$. The solid
curves are our theoretical expectation
from Eq.\ (\ref{surmise}), the data points 
result from the numerical solution of the BdG equation.
The inset shows the numerical data for the
mean density of states.
}
\label{fig:SNS}
\end{figure}

In Fig.\ {\ref{fig:SNS}} we compare Eq.\ (\ref{surmise})
with numerical data. The agreement is excellent for a range of
mean free paths in the quasiballistic regime. The mean 
position of the splitted levels fluctuates only to
higher orders in $L/\ell$. 
This makes it possible to resolve the splitting in the mean density of
states (see inset in Fig.\ {\ref{fig:SNS}}).
\vfill

\section{Localized regime}

\subsection{NS junction}

In order to go beyond  the quasiballistic regime
into the localized regime $L\gg \ell$ we
write the quantization condition for the Andreev levels
in an NS junction in the form
\be
r(\ep)r(-\ep)^* e^{2 i\phi_A}=1,
\e
where $r(\ep)=e^{i\phi(\ep)}$ is the reflection amplitude
of the disordered normal metal.
[The hole has reflection amplitude $r^*(-\ep)$.]
In terms of the phase shifts we have
\be
\Phi(\ep)\equiv \frac{\phi(\ep)-\phi(-\ep)}{2} +\phi_A(\ep)=\pi n,
\; n=0,1, \dots.
\e
The density of states $\nu(\ep)=\sum_n\delta(\ep-\ep_n)$ 
is related to the scattering phase shifts by{\cite{DS}}
\be
\label{nodecounting}
\nu(\ep)=-{1\over \pi} \frac{d}{d\ep} \im \ln \sin
\Phi(\ep+i 0^+),
\e
where $0^+$ denotes a positive infinitesimal.
The imaginary part of the logarithm
jumps by $\pi$ whenever $\sin\Phi(\ep)$
changes sign, hence it counts the number of 
levels below $\ep$.
The derivative with respect to $\ep$
then gives the density of states.
It is convenient to write Eq.\ (\ref{nodecounting})
as a Taylor series,
\be
\label{series}
\nu(\ep)={1\over \pi}\frac{d}{d\ep}
\l(\Phi+\im \s_{m=1}^\infty
{1\over m} e^{2i m\Phi} \r),
\e
which converges because $\Phi(\ep+i0^+)$
is equivalent to $\Phi(\ep)+ i0^+$.

We seek the disorder averaged density of states 
$\la \nu(\ep)\ra$. One way to proceed is 
by means of the Berezinskii technique.\cite{Berezinskii,BG}
An alternative way, that we will follow here, is to start
from the scaling equation\cite{White,TB} for the probability distribution
$P(\phi_N)$ of the phase shift 
$\phi_N=\case{1}{2}[\phi(\ep)-\phi(-\ep)]$.
This equation has the form
\be
\label{P}
{\pa P\over \pa L}={\pa\over \pa \phi_N}
\l(-{2 \ep\over \hbar v_F}+{1\over \ell}
{\pa\over \pa \phi_N}\sin^2{\phi_N}
\r) P.
\e
The initial condition is $\lim_{L\to 0} P(\phi_N)=\delta(\phi_N)$.

The first moment satisfies 
${\pa\la \phi_N \ra / \pa L}={2 \ep/ \hbar v_F}$,
hence
\be
\label{meanphi}
\la \phi_N \ra={2 \ep L\over \hbar v_F}.
\e
Multiplication of
Eq.\ (\ref{P}) by $\exp{(2im\phi_N)}$
and integration over $\phi_N$ from $0$ to $\pi$
yields a set of recursive differential equations\cite{Berezinskii}
for the moments 
$R_m=\la e^{2 i m \phi_N}\ra$,
\be
\label{Rm}
{\pa R_m\over \pa L}=
{m^2\over \ell}(R_{m+1}+R_{m-1}-2 R_m)+{4 i \ep \over \hbar v_F}m R_m,
\e
with the initial condition $R_m(0)=1$. We solve this set of equations
by truncating the vector $(R_1, R_2, \dots R_M)$
at a sufficiently large value of $M \approx 400$ and diagonalizing
the corresponding tri-diagonal matrix. From Eq.\ (\ref{series})
we then find the mean density of states.

\begin{figure}
\hspace*{3cm}
\epsfxsize10cm
\epsfbox{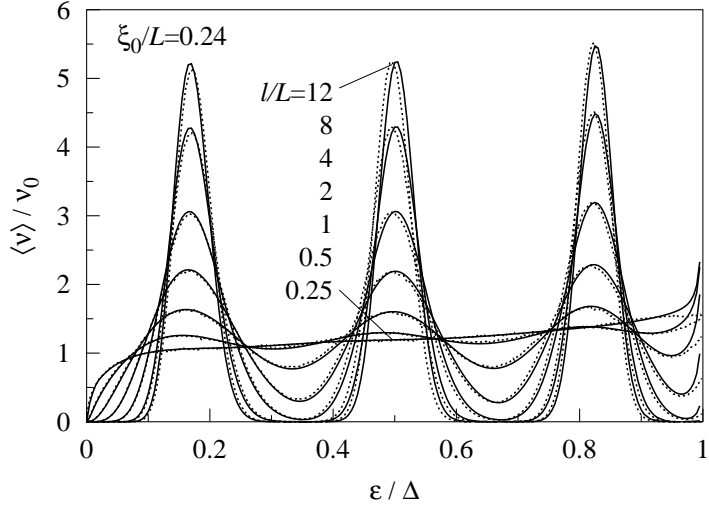}
\medskip\\
\caption{Mean density of states 
of an NS junction from the quasiballistic into the
localized regime. The solid curves have been computed from
Eqs.\ (\ref{series}) and (\ref{Rm}).
The dashed curves are
a numerical simulation
of random disorder in the BdG
equation.}
\label{fig:Rm}
\end{figure}

The result is shown in Fig.\ {\ref{fig:Rm}} for 
$\xi_0/L=0.24$ and ratios $\ell/L$
ranging from the quasiballistic regime to the localized regime.
Agreement with the numerical solution of the BdG equation
is excellent over the whole range.

In the localized  regime $L\gg \ell$ the individual
Andreev levels can no longer be distinguished
in the mean density of states, because the broadening
of the levels becomes greater than the spacing.
In this regime we distinguish two
energy ranges, $\ep \gg \ep_g$
and $\ep \ll \ep_g$,
where $\ep_g=(\hbar v_F/\ell)e^{-L/\ell}$.

For energies higher than $\ep_g$ we may 
use the $L\to \infty$ limit of the distribution
$P(\phi_N)$, obtained by setting the 
left-hand-side of Eq.\ (\ref{P}) equal to zero.
The resulting moments are
\be
\lim_{L\to\infty}R_m=\i_0^\infty d\sigma\;
e^{-\sigma} \l( {\sigma\over \sigma -i \omega}\r)^m,
\qquad \omega={4\ep \ell\over \hbar v_F}.
\e
We then calculate the mean density 
of states from Eq.\ (\ref{series}), with the result
\beq
\label{flat}
&&\la \nu(\ep)\ra={2 L\over \pi \hbar v_F}
+{1\over \pi \sqrt{\Delta_0^2-\ep^2}} + f(\ep),\quad \ep\gg \ep_g,\\
&&f(\ep)=\!{\pa\over \pa\ep} \im \!\!
\i_0^\infty{d\sigma\over\pi}\l[\frac{e^{-\sigma}}{\sigma\!-\! i\omega}-
\frac{e^{-\sigma}(1-e^{2i\phi_A})}{\sigma(1-e^{2i\phi_A})\!-\! i\omega}\r].
\eq
The first term on the right-hand-side of
Eq.\ (\ref{flat}) is the energy independent density 
of states $\nu_0$ in an isolated normal metal.
The main effect of the superconductor 
for $\ep\gg \ep_g$ is an
enhancement of the density 
of states close to the gap
$\Delta_0$ of the bulk superconductor (second term).
The third term is negative for sufficiently small 
$\ep$ and is a precursor of the soft
gap near the Fermi level.
For $\xi_0\ll \ell$ and $\ep\ll \hbar v_F/\ell$
the reduction term $f(\ep)$
can be simplified as 
\be
f(\ep)=-{2\ell \over \pi \hbar v_F}
\l(\ln \frac{\hbar v_F}{8 \ep \ell}-\gamma \r),
\qquad \ep_g\ll \ep \ll {\hbar v_F\over \ell},
\e
where $\gamma\approx 0.58$ is Euler's constant. 

Near the Fermi level, for $\ep\ll \ep_g$,
the mean density of states vanishes as a result of the proximity 
to the superconductor. This ``soft gap'' appears no matter how 
strongly localized the normal metal is.
The coefficients $R_m$ may now be treated as
analytical functions of the parameter 
\be
z=-{4 i \ep \ell m\over \hbar v_F}, \qquad R_m=R(z).
\e 
Taking the limit $\ep \to 0$ we deduce from
Eq.\ (\ref{P}) the partial differential equation  
\be
\label{basic}
\ell {\pa R\over \pa L}=z^2 {\pa^2 R\over \pa z^2}
-z R,
\e
with initial condition $\lim_{L\to 0}R(z)=1$. 
This differential equation has been studied before
in the theory of one-dimensional localization,{\cite{AP,FM}}
but not in connection with the proximity effect.
The result for the mean density of states,
derived in the Appendix, is given by
\be
\label{theend}
\la \nu(\ep) \ra=\frac{2 \ell}{\pi^{3/2} \hbar v_F}
\exp{\l[-\frac{\ell}{4 L}\ln^2{\pi \ep_g\over \ep}
-{u\ell\over 2L}\l(\ln{u \ell \over 2 L}-1\r) \r]},
\e
where $u=\ln \pi \hbar v_F/\ep \ell=\ln\pi\ep_g/\ep +L/\ell$.
The leading logarithmic asymptotic of this expression 
in the limit $\ep\ll \ep_g$ has the log-normal tail 
\be
\label{nu4}
\la \nu(\ep) \ra\propto
\exp{\l[-\frac{\ell}{4 L}\ln^2{\pi \ep_g\over \ep}\r]},
\qquad \ep\ll \ep_g.
\e

The same log-normal tail was found in Ref.\ {\onlinecite{Skworzov}}
for a many-channel diffusive conductor. In that case the factor 
$\ell/L$ is replaced by the Drude conductance of the normal metal
and the energy scale $\ep_g$ is replaced by the Thouless energy 
$\hbar D/L^2$ (with $D$ the diffusion constant).
In our single-channel localized conductor nether the
Drude conductance nor the Thouless energy play a role.

\subsection{SNS junction}

\begin{figure}
\hspace*{3cm}
\epsfxsize10cm
\epsfbox{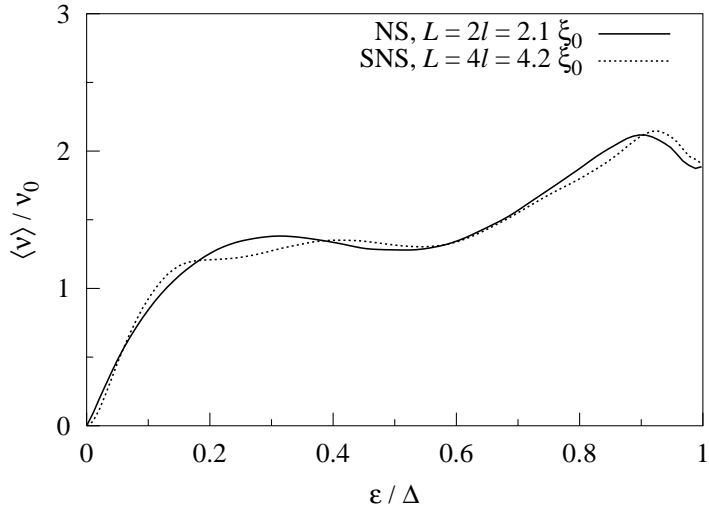}
\medskip\\
\caption{
Numerical calculation of the mean density of states 
of an NS junction (solid) and SNS junction (dashed)
in the nearly localized regime. The length of the SNS
junction is twice that of the NS junction. 
(The weak oscillations are remnants of Andreev levels,
that will disappear if $L/\ell$ is increased further.)
}
\label{fig:local}
\end{figure}

In contrast to the quasiballistic regime, the NS and SNS junctions
are similar in the localized regime. (At least for the case of zero current
through the SNS junction considered here.) Unfortunately, there exists no
simple scaling equation as Eq.\ (\ref{P}) that can describe 
the density of states of the SNS junction. We therefore rely on the numerical solution
of the BdG equation. In Fig.\ {\ref{fig:local}} we show that 
the mean density of states of an NS junction of length $L$
is close to that of an SNS junction of length $2L$.
This factor of $2$ has an obvious explanation in the ballistic regime
[compare Eqs.\ (\ref{NSlevels}) and (\ref{SNSlevels})], but it is remarkable 
that it still applies to the localized regime.
\vfill

\section{Conclusion}

In summary, we have calculated the effect of disorder on the spectrum
of Andreev levels in single-channel NS and SNS junctions.
The non-perturbative effects of localization in the one-dimensional case
can be studied exactly, at least in the NS geometry.
Our research is of theoretical interest in view of
recent studies of the sub-gap density of states
beyond mean field theory,\cite{Beloborodov,Simons,Vavilov,Skworzov}
but may also be of experimental interest in view of recent progress made 
in superconductor--carbon-nanotube devices.\cite{MKMD,Wei} 

\acknowledgements

We thank Piet Brouwer for a crucial discussion
at the initial stage of this project.
This research was supported by the 
``Ne\-der\-land\-se or\-ga\-ni\-sa\-tie
voor We\-ten\-schap\-pe\-lijk On\-der\-zoek'' (NWO) and by the
``Stich\-ting voor Fun\-da\-men\-teel On\-der\-zoek der Ma\-te\-rie'' 
(FOM). 
M.T. and N.A.M. thank the visitors program at the Max-Planck-Institut
f\"{u}r Physik komplexer Systeme, Dresden. N.A.M. also acknowledges
support by the ``Ingeni\o rvidenskabelig Fond og G.A. Hagemanns
Mindefond''.

\appendix
\section{Derivation of the log-normal tail}

The differential operator on the right-hand-side of
Eq.\ ({\ref{basic}}) has eigenfunctions
\be
\label{function}
f_p(z)=2\sqrt{z}K_p(2\sqrt{z}), 
\e
where $K_p(z)$ is the modified Bessel function,
such that
\be
\l(z^2 {\pa^2\over \pa z^2} -z\r) f_p(z)=\frac{p^2-1}{4}f_p(z).
\e
The solution to Eq.\ ({\ref{basic}}) with the
initial condition $\lim_{L\to 0}R(z)=1$ is 
\be
\label{solutionR}
R(z)=f_1(z)+\!\!\i_{-\infty}^\infty\!\! d\nu\,
\frac{\nu \sinh(\pi\nu/2)}{\pi(\nu^2+1)} f_{i\nu}(z)
e^{-(\nu^2+1)L/4\ell}.
\e

To obtain the density of states of the NS junction it is convenient
to define the inverse Laplace transform
\be
\label{inverse}
F(\lambda)={1\over 2\pi i} \i_{-i\infty+0^+}^{i\infty+0^+}{dz\over (4\lambda)^2}R(z) 
\exp{\l({z\over 4\lambda}\r)}.
\e
From Eq.\ (\ref{series}) we find for $\ep \ll \ep_g$
the mean density of states in terms of the function $F$,
\be
\label{relation}
\la \nu(\ep) \ra={4 \ell \over \pi \hbar v_F } F\l({\ep \ell\over \pi \hbar v_F}\r).
\e

Our aim is to find the asymptotic form of $F(\lambda)$ in
the limit $\lambda\to 0$.
The inverse Laplace transform of the modified Bessel functions 
in Eq.\ (\ref{solutionR}) can be found in Ref. \onlinecite{PBM}.
We obtain
\beq
\label{F}
&&F(\lambda)=F_0(\lambda)-\i_{-\infty}^{\infty}d\nu\,
\lambda^{-(i\nu+1)/2} e^{-(\nu^2+1)L/4\ell }\n \\
&&\qquad
\times{\,_1\!F_1\l(\case{3}{2}+\case{i\nu}{2},1+i\nu,-4\lambda \r) 
\over 2\sqrt{\pi}\,(1-i\nu)\Gamma\l(i\nu/2\r)}, 
\eq
where $F_0(\lambda)=\exp(-4\lambda)$.
The integrand has a single pole $\nu=-i$
in the lower half of the complex plane
and the residue from this pole cancels the term $F_0$.
Let us shift the contour by the transformation
$\nu\to \nu -({i\ell/L})\ln{(1/\lambda)}$
and consider the limit $\lambda\ll e^{-L/\ell}$.
In this limit the contour is shifted through
the pole so that the term $F_0$ is cancelled.
Moreover, the hypergeometric function $\,_1\!F_1$ 
can be replaced by unit in this limit.
Thus, we end up with the integral
\beq
\label{F1}
&&F(\lambda)={1\over 2\sqrt{\pi}}
e^{-\case{\ell}{4L}\l(\ln{\case{1}{\lambda}}-\case{L}{\ell} \r)^2}
\i_{-\infty}^{\infty}d\nu\,
e^{-\nu^2 L/4\ell}\n \\
&& \qquad
\times\l[\l(i\nu-1-{\ell\over L}\ln{\lambda}\r)
\Gamma\l({i\nu\over 2}-\frac{\ell}{2 L}\ln{\lambda}\r)\r]^{-1}.
\eq
The asymptotic form of this integral in the limit $\lambda\ll e^{-L/\ell}$
can be found by evaluation of the expression in square brackets 
in the point $\nu=0$ and calculation of the 
Gaussian integral. 
Using the asymptotic formula for the Euler gamma
function one obtains the mean density of states
given in Eq.\ (\ref{theend}).


\begin{references}

\bibitem{Beloborodov}
I. S. Beloborodov, B. N. Narozhny, and I. L. Aleiner,
Phys.\ Rev.\ Lett. {\bf 85}, 816 (2000).

\bibitem{Simons}
A. Lamakraft and B. D. Simons, 
Phys.\ Rev.\ Lett. {\bf 85}, 4783 (2000);
cond-mat/0101080.

\bibitem{Vavilov}
M. G. Vavilov, P. W. Brouwer, V. Ambegaokar, and C. W. J. Beenakker,
Phys.\ Rev.\ Lett. {\bf 86}, 874 (2001).

\bibitem{Skworzov}
P. M. Ostrovsky, M. A. Skvortsov, and M. V. Feigel'man,
cond-mat/0012478. 

\bibitem{Andreev} 
A. F. Andreev, Sov.\ Phys.\ JETP {\bf 19}, 1228 (1964).

\bibitem{CWJreview}
C. W. J. Beenakker,
Rev.\ Mod.\ Phys. {\bf 69}, 731 (1997).

\bibitem{White} 
B. White, P. Sheng, Z. Q. Zhang, and G. Papanicolaou, Phys.\
Rev.\ Lett. {\bf 59}, 1918 (1987).

\bibitem{TB}
M. Titov and C. W. J. Beenakker, 
Phys.\ Rev.\ Lett. {\bf 85}, 3388  (2000). 

\bibitem{TBMF}
M. Titov, P. W. Brouwer, A. Furusaki, and C. Mudry,
cond-mat/0011146. 

\bibitem{Mehta} 
M. L. Mehta, {\it Random Matrices}
(Academic, New York, 1991). 

\bibitem{Mirlin}
A. D. Mirlin, 
Phys.\ Rep. {\bf 326}, 259 (2000).

\bibitem{deGennes}
P. G. de Gennes, {\it Superconductivity  of Metals and Alloys}
(Benjamin, New York, 1966).

\bibitem{DS} 
E. Doron and U. Smilansky, 
Phys.\ Rev.\ Lett. {\bf 68}, 1255 (1992).

\bibitem{Berezinskii} 
V. L. Berezinskii,
Sov.\ Phys.\ JETP {\bf 38}, 620 (1974).

\bibitem{BG}
V. L. Berezinskii and L. P. Gor'kov,
Sov.\ Phys.\ JETP {\bf 50}, 1209 (1979).

\bibitem{AP} 
B. L. Altshuler and V. N. Prigodin,
Sov.\ Phys.\ JETP {\bf 68}, 198 (1989).

\bibitem{FM}
Y. V. Fyodorov and A. D. Mirlin,
Int.\ J.\ Mod.\ Phys.\ B {\bf 8}, 3795 (1994).

\bibitem{PBM}
A. P. Prudnikov, Yu. A. Brychkov, and O. I. Marichev,
{\it Integrals and Series}
(Gordon and Breach,  Amsterdam, 1986).

\bibitem{MKMD}
A. F. Morpurgo, J. Kong, C. M. Marcus,
and H. Dai, Science {\bf 286}, 263 (1999).

\bibitem{Wei}
Y. Wei, J. Wang, H. Guo, H. Mehrez, and C. Roland,
condmat/0103210.

\end{references}
\end{document}